# Detection of the third innermost radiation belt on LEO CORONAS-Photon satellite around 2009 solar minimum

Oleksiy V. Dudnik[a], Janusz Sylwester[b], Mirosław Kowaliński[b], Piotr Podgórski[b], Kenneth J. H. Phillips[c]

[a]Institute of Radio Astronomy, National Academy of Sciences of Ukraine, Mystetstv St. 4, Kharkiv, 61002, Ukraine
[b]Solar Physics Division, Space Research Centre, Polish Academy of Sciences, Poland
[c]Scientific Associate, Dept. Earth Sciences, Natural History Museum, London SW7 5BD, United Kingdom

**Abstract**

We analyze variations of high energy charged particle populations filling various magnetospheric regions under, inside and outside of the Van Allen inner and outer electron radiation belts in May 2009. The study is based on the experimental data obtained from the STEP-F and the SphinX instruments placed close to each other aboard the low-Earth circular orbit CORONAS-Photon satellite. Data analysis of particle fluencies collected from the highly sensitive STEP-F device indicates the presence of a persistent electron belt at $L \approx 1.6$, i.e., beneath the well-known Van Allen electron inner radiation belt of the Earth's magnetosphere. The electron energy spectrum in this "new" belt is much steeper than that of the inner belt, so that the electrons with energies $E_e \geq 400$ keV were almost not recorded on $L \approx 1.6$ outside the South Atlantic Anomaly (SAA). We introduce the concept of effective lowest threshold energies for X-ray detectors used in the solar soft X-ray spectrophotometer SphinX and define their values for two regions: the SAA and in the Van Allen outer belt. Different values of lowest threshold energies are directly associated with different slopes of particle energy spectra. Cross-analyses of data obtained from the STEP-F and SphinX instruments initially built for various purposes made it possible to detect the highly anisotropic character of the spatial electron distribution in radiation belts in both Southern and Northern hemispheres. We detected also the presence of low-energy electrons at all latitudes during the main phase of a weak geomagnetic storm.

1. Introduction

Populations of electron radiation belts filled with relativistic electrons often show profound dynamic changes during intense geomagnetic storms and substorms. These variations are associated with different physical mechanisms that include transport, acceleration and loss processes. Frequently observed relativistic electron precipitation in the radiation belt at McIlwain parameter $L \approx 4 - 6$ with abruptly generating temporal structure lasting less than a few seconds were observed by the Solar Anomalous and Magnetospheric Particle Explorer (SAMPEX) spacecraft [Blake and Selesnick, 2001], and were actively studied in recent decades [Imhof et al., 1992; Nakamura et al., 1995, 2000; Blake et al., 1996; Lorentzen et al., 2001; Baker et al., 2019]. Currently the research interest is focused on the investigation of bursty regions of the Van Allen inner radiation belt at the heights of low-Earth orbit satellites in middle and low geographic latitudes, in regions far from the South Atlantic Anomaly (SAA). The *L* - dependent peaks in the quasi-trapped electron populations have been observed over a broad geographical longitude range with a preference for occurring in the hours close to local dawn and dusk [Imhof et al., 1978]. Spiky-type spectra were observed progressively more frequently at longitudes further East of the SAA. Over a period of a year, the flux peak at $L = 1.65$ had been found to undergo long-term changes [Imhof et al., 1978].

---

*Corresponding author at: Department of Space Radio Physics, Institute of Radio Astronomy, National Academy of Sciences of Ukraine, 61002 Mystetstv str., 4, Ukraine.
E-mail addresses: dudnik@rian.kharkov.ua; o.v.dudnik@nas.gov.ua. Phone: +380 506420015

A precipitation of high energy protons and electrons was observed at $L = 1.61$ during March 1984 - September 1986 on the Japanese OHZORA satellite [Nagata et al., 1988]. It was concluded that not only electron (0.19 - 3.2 MeV) but also proton (0.58 – 35 MeV) precipitating fluxes were observed at low latitudes; energies of registered particles were within an interval of one-MeV; the highest fluence peak location of precipitating electrons and protons has been observed at $L = 1.71$; the width of the $L$-peak was $\Delta L = \pm 0.16$. The geographical distribution showed no substantial particle precipitation over the Northern Hemisphere zones that are the conjugate points of the SAA. Also, no positive correlation was found with $K_p$-index of the geomagnetic activity [Nagata et al., 1988]. A summary map of distribution of electron fluxes at a height of ~ 600 km shows that particle fluxes were not distributed uniformly; three specific enhanced density zones have been observed depending on latitude: the equatorial zone ($L < 1.2$); the low latitude ($1.2 < L < 1.4$); and the middle latitude zone ($1.6 < L < 2$) [Grachev et al., 2005; Grigoryan et al., 2008a, b; Sadovnichy et al., 2011]. The main reason why the authors separated zones near the equator and in the middle latitudes was that flux splashes were observed sporadically on the equator; meanwhile, such fluxes were detected almost continuously in the middle latitudes.

The aforementioned investigations do not explain unambiguously the mechanisms of sporadically occurring splashes of electron counts in the middle and low latitudes. One of the proposed mechanisms is energetic electron flux enhancement in the drift loss cone associated with lightning strikes [Blake et al., 2001; Rodger et al., 2003; Martinez-Calderon et al., 2020]. SAMPEX and Van Allen Probes (formerly known as Radiation Belt Storm Probes or RBSP) data have revealed many cases of drift loss cone flux enhancements in localized $L$-shell regions, indicating that the precipitation of electrons with energies of several hundred keV into the drift loss cone is commonplace. This conclusion was made during the time intervals when the satellites are properly positioned to detect electrons in the drift loss cone only. This precipitation is apparently driven by whistler waves originating in thunderstorms active around the globe at any given time. Sporadically generated and continuously existing additional radiation belt(s) below the well-known Van Allen outer electron belt were observed repeatedly [Blake et al., 1992; Blake and Selesnick, 2001; Hudson et al., 2008; Baker et al., 2013], while references to similar formations beneath the Van Allen inner belt are almost absent.

In this work we analyze enhanced flux electron volumes outside the Van Allen belts observed during the minimum of the eleven-year solar activity cycle based on data obtained from two space-borne scientific instruments – the Satellite Telescope of Electrons and Protons (STEP-F) and the Solar PHotometer IN X-rays (SphinX) on board the CORONAS-Photon spacecraft, which were a part of the science payload package. CORONAS-Photon was launched on 2009 January 30 into a near-polar, low-Earth orbit (inclination 82.5 degrees, altitude 550 × 550 km) and operated until 2009 December when power failures terminated the mission.

The instruments had orthogonally oriented viewing axes relatively to spacecraft's OZ axis with SphinX pointing towards the center of the solar disc. The aim of STEP-F was to monitor electron, proton and alpha-particle fluxes and spectra in a number of energy ranges. Though the SphinX was oriented continuously at the solar disc and energetic spectra of coronal soft X-ray emission were measured, the two most energetic channels of the 256-channel energy spectrum contained information about the Earth's magnetosphere energetic charged particles interacting with detectors and/or mechanical parts near the detectors. We took advantage of using SphinX data to study the energetic charged particles in the magnetosphere following previous work which used space-borne and balloon X-ray and gamma-ray detectors to study terrestrial radiation belts [Imhof et al., 1985; Arkhagelskaja et al., 2008; Millan et al., 2011, 2013; Woodger et al., 2015]. Evidence presented here indicates for the first time the presence of a new permanent radiation belt beneath the Van Allen inner belt.

## 2. Brief description of space instruments used for analysis

### 2.1. The satellite instrument STEP-F.

The STEP-F electron and proton telescope was designed to study distributions and properties of high energy charged particles within and beneath the Earth radiation belts. The instrument was part of the science payload of the CORONAS−Photon satellite [Dudnik et al., 2011; Kotov, 2011]. The STEP-F was included in the STEP-FD unit, attached to the outer surface of the spacecraft, and to the digital electronic unit STEP-FE located inside the sealed pressurized chamber [Dudnik et al., 2019]. The detector head of the STEP-FD unit, constructed as a collimating system, contained two silicon position-sensitive detectors, namely D1 and D2, and two single-crystal

CsI(Tl) scintillators, namely D3 and D4. The D3 scintillator was viewed by large-area silicon PIN photodiodes, and D4 was viewed by a vacuum photomultiplier tube [Kudin et al., 2010]. The bottom scintillation detector D4 in the telescope system of the STEP-FD unit detected secondary gamma quanta generated as a result of the interaction of electrons with the upper layers of detectors and construction materials.

The angle of view of STEP-F changed from 108° × 108° to 98° × 98° depending on particle energy. Active areas of both upper silicon detectors D1 and D2 were $S_1 = S_2 = 17$ cm$^2$, and of CsI(Tl) detectors D3 and D4 were $S_3 = 36$ cm$^2$ and $S_4 = 49$ cm$^2$, respectively. Table 1 gives the energy channels of registered electrons and protons and their titles. A description and characteristics of STEP-F's and SphinX's channels used in analysis are given by Dudnik et al., 2014.

Table 1. Energy ranges of registered electrons and protons.

| **Particle sort** | **Energy range, keV** | **Channel** | **Remark** |
|---|---|---|---|
| electrons | 0.35 – 0.95<br>1.2 – 2.3<br>> 2.3 | D2e<br>D3e1<br>D3e2 | Provided by D2 silicon matrix detector and two energy channels of D3 CsI(Tl) scintillation detector |
| protons | 7.4 – 10.0;<br>15.6 – 55.2 in 9 channels;<br>> 55.2 | D2p<br>D3p(1-9)<br>D3p10 | Provided by D2 silicon matrix detector and 10 energy channels of D3 CsI(Tl) scintillation detector |
| electrons + protons | 0.18 - 0.51 (electrons)<br>+ 3.5 - 3.7 (protons) | D1e | Channel of mixed particle registration; provided by D1 silicon matrix detector |
| electrons + protons | 0.55 – 0.95 (electrons)<br>+ 3.7-7.4 (protons) | D1p | Channel of mixed particle registration; provided by D1 silicon matrix detector |
| bremsstrahlung | see in the text | D4e | Provided by D4 CsI(Tl) scintillation detector |

*2.2. The X-ray spectrophotometer SphinX*

The SphinX instrument was developed at the Solar Physics Division of the Space Research Centre, Polish Academy of Sciences [Sylwester et al., 2008, Gburek et al., 2011a]. Like STEP-F, SphinX was part of the science payload of the CORONAS−Photon satellite [Gburek et al., 2011b]. Four circular silicon PIN photodiodes of thickness 500 μm were used to detect X-rays with energies lower than 15 keV. However, the crystal sensitive areas differed from each other. On the sunward side, 12.5 μm beryllium foils were mounted, as well as a 320 nm aluminized Mylar foil used to eliminate optical and EUV radiation. The sensor with an active area of $S_2 = 0.111$ cm$^2$ (SDet2) was covered by tantalum plate with a thickness of 400 μm while maintaining an open window of $S_{2o} = 4.9 \times 10^{-3}$ cm$^2$ for detection of X-rays from powerful solar flares. The entire detection range $\Delta E = 0 – 15$ keV was distributed over 256 energy channels of the spectrometer analog-to-digital converter (ADC). The energy threshold of signal detection of Detector 1 (SDet1) sensor was set to 1.2 keV, that of Detector 2 (SDet2) detector to 0.85 keV. The axes of viewing cones of the three detectors were directed along the OZ axis of the spacecraft, which was usually pointed toward the center of the solar disc. The fields of view of SDet1 and SDet2 were 1.94° × 1.94° and 1.72° × 1.72°, respectively, centered along axes.

The signals were recorded in 256 amplitude channels, corresponding to different energies of incoming solar photons in the X-ray band. However, we noticed that the counting rates recorded by detectors increased substantially during passages through the Van Allen belts and the SAA. The largest increase was observed for the last, highest energy channel of the SDet1 and SDet2 detectors. Thus, we assumed that high-energy charged particles were also registered in the highest ADC channels. The data from 255[th] energy channels of SDet1 and SDet2 sensors were used for the analysis – detections in these channels have energies above about 15 keV.

## 3. Evidence of three electron radiation belts as seen with STEP-F

The specific feature of CORONAS-Photon satellite's orbit was its crossing of the same near-Earth space environment every 15$^{th}$ orbit pass, or approximately every 24 hours. As particle distributions inside the Earth's magnetosphere depend on the latitude, longitude and the height, it is useful to compare fluxes detected at the same or similar spatial locations at consecutive time intervals. As the satellite's orbit was very close to circular, these requirements are restricted by the latitude and longitude only. The longitude drift of the orbit constituted less than 2.5 degrees between starting and following (+15) orbit passes at a particular latitude. By combining fluencies recorded over the same geographic area from several passes we can trace the temporal variability of electron fluxes. Figure 1 demonstrates the time dependence of electron flux density with energies $\Delta E_e = 0.18$–$0.51$ MeV derived from the D1e energy channel data with 30-second time resolution over the period from 1$^{st}$ till 31$^{st}$ May 2009 as recorded on every ninth orbit pass of the day (Universal Time). From bottom to top, the signal variations are shown from the Southern to Northern hemispheres. It is worth noting that the ninth orbits of the day pass far from the SAA zone.

A cursory look at Figure 1 reveals the presence of a third radiation belt in addition to the well-known magnetospheric belts which were crossed by the satellite at 35 to 32 degrees south, and at latitudes approximately 36 to 42 degrees north. These latitudes correspond to $L \approx 1.6$. Satellite crossings of the Van Allen radiation belts ($L \approx 2.3$ and $L \approx 4$ to 5) are clearly pronounced in both hemispheres. Figure 1 shows enhancement of particle flux in all three belts in the vicinity of the weak geomagnetic storm ($D_{st} = -25$ nT) of May 8, 2009. It is seen also a gradually increasing trend of latitude with maximum particle intensity versus day of the month. This trend is explained by the longitudinal drift of orbit's projection onto the Earth's surface at fixed geographic latitude as can be seen in Figure 2. In this Figure, a shaded blue strip represents the longitudinal drift of the CORONAS-Photon satellite from 1 to 31 May 2009. A strip is pasted in front of the global map of electron fluxes in the energy range $\Delta E_e = 0.6 - 1.2$ MeV derived by the space radiation detector SDOM aboard the polar-orbiting satellite Daichi (Advanced Land Observing Satellite or ALOS) at the altitude of 700 km [Obara et al., 2012]. The longitudinal drift leads to change of geomagnetic latitudes. As a result, particle fluxes are followed by shifted geomagnetic latitudes.

It was stated by Dudnik, 2010 that two inner radiation belts at $L \approx 2.28$, and at $L \approx 1.61$ were detected in the energy range $\Delta E_e = 0.18 - 0.51$ MeV. The persistent presence of the third belt on lower $L$-shell, and the time variations of the electron population in both belts are observed to depend on the overall level of geomagnetic activity.

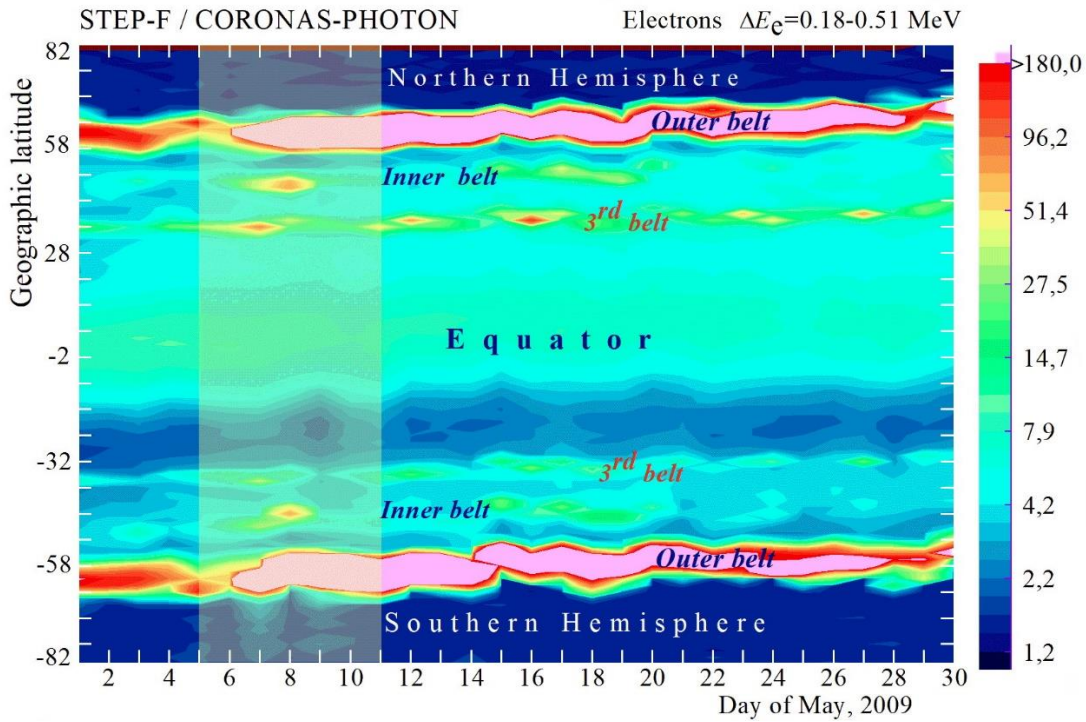

Figure 1. Pattern of evolution of derived electron fluxes as obtained from the STEP-F measurements in May 2009. The color scale on the right of the figure shows electron flux densities in units of particles × (cm$^2$·s·sr)$^{-1}$ for each consecutive with 30-second time resolution along the orbit. The shaded area indicates the substorm time interval.

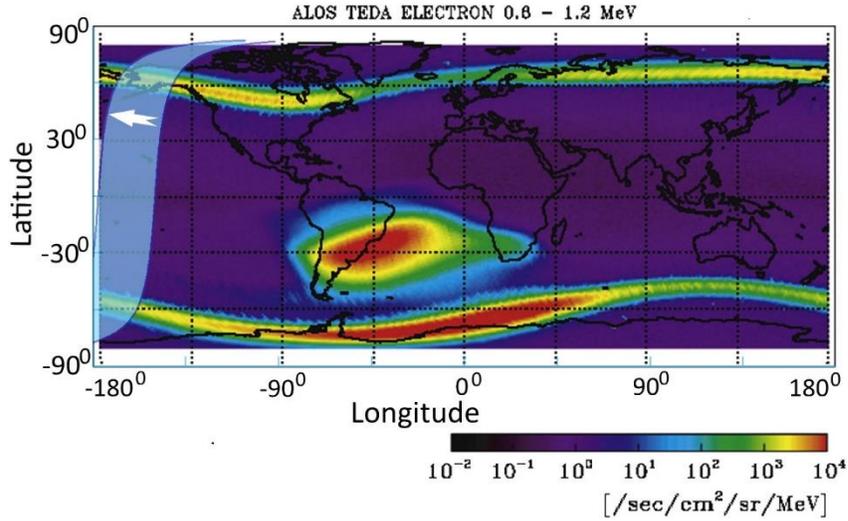

Figure 2. The longitudinal drift of CORONAS-Photon orbit's projection (marked by the curved blue strip arrowed) onto the global map of electron fluxes as measured by Daichi (ALOS) spacecraft. The colored horizontal scale at the footnote of the figure shows the electron fluxes measured by Daichi [Obara et al., 2012].

All belts are clearly seen at geographical longitudes away from the SAA. Figure 3 shows the time variations of the particle flux density in the Van Allen inner and the third radiation belts at their flux density maxima in the two energy ranges on the ninth diurnal ascending parts of the orbits in the Northern hemisphere throughout the entire month. The $K_p$-index is shown at the top panel to display a level of geomagnetic activity. The data are presented on a logarithmic scale in the energy range $\Delta E_e$ = 0.18 to 0.51 MeV and on a linear scale in the energy range $\Delta E_e$ = 0.35 to 0.95 MeV. It can be seen from Figure 3 that the ratio of electron flux density observed in the third and Van Allen inner belts (plotted by a dash and solid lines respectively) exceed 1 in 25 out of 31 days in the energy range $\Delta E_e$ = 0.18 to 0.51 MeV, and the respective ratio is less than 1 in 26 out of 31 days for electrons with energies $\Delta E_e$ = 0.35 to 0.95 MeV. Hence, the energy spectra of electrons populating both innermost belts differ significantly: a harder spectrum was recorded in the "classical" inner Van Allen belt, while the energy spectrum in a newly detected third one is definitely softer.

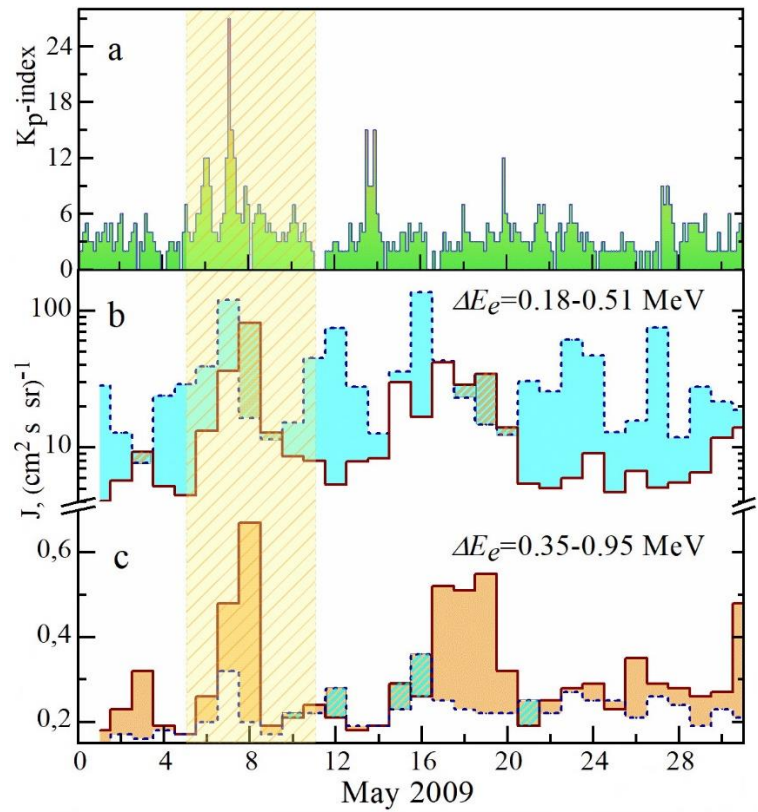

Figure 3. $K_p$-index (**a**), particle flux density in the Van Allen inner (dashed) and the third radiation belts in two energy bands (**b**,**c**) throughout May 2009. The shaded area corresponds to the substorm time interval.

The geophysical conditions in the first half of May, 2009 are shown in Figure 4. The main parameters of space weather are presented, including the planetary geomagnetic and near-equatorial $K_p$ - (**a**) and $D_{st}$ - indices (**b**), velocity $V_{sw}$ (**c**) and density $n$ (**d**) of the solar wind as measured by the proton monitor Charge, Element, and Isotope Analysis System (CELIAS/MTOF/PM) on the Solar and Heliospheric Observatory (SOHO) spacecraft. Figure 4 also shows fluxes of electrons with energies $E > 2$ MeV (**e**) and protons with energies $E_p = 0.8 – 4.0$ MeV (**f**) as observed from the GOES-11 geostationary satellite. In May 2009 the particle fluxes at outer edge of the outer belt were measured simultaneously by two other satellites: GOES-10, and GOES-12. All three observed similar trends. This is why in Figure 4 we show results of measurements from only a single representative GOES satellite. From Figure 4, it is seen that within the period 6 to 14 May, a weak geomagnetic storm ($D_{st} \approx$ - 30 nT) took place indeed with the main phase on May 8 starting from ~ 06:00 till ~ 09:00 UT. The initial stage of the storm was from ~ 06:00 till ~ 09:00 observed at the beginning May 6 and was not accompanied by a sudden commencement. At the same time, during the initial stage a local enhancement of magnetic field indices occurred coinciding with the maximum gradient of a solar wind enhancement observed close to midnight between May 6 and 7. Immediately after a local growth of the solar wind density during initial stage of the storm on May 6 and 7 STEP-F observed a significant increase of electron fluxes in both the Van Allen inner radiation belt and the third electron belt mentioned above (see Figure 5).

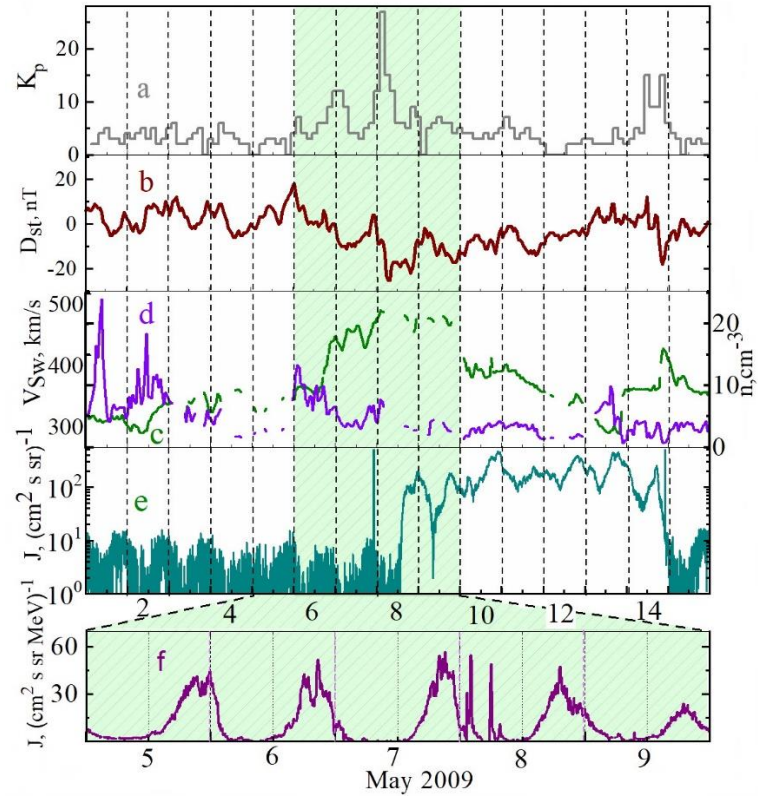

Figure 4. Changes of important parameters characterizing geomagnetic activity associated with the geomagnetic substorm of May 8, 2009: $K_p$ (**a**), and $D_{st}$ (**b**) -indices; $V_{sw}$ (**c**) and $n$ (**d**) of solar wind; electron (**e**) and proton (**f**) fluxes at geostationary orbit.

Figure 5 shows particle flux dynamics in two energy ranges. Plotted are electron fluxes with energies $\Delta E_e$ = 0.18 - 0.51 MeV (**a**), and with energies $\Delta E_e$ = 0.35 - 0.95 MeV (**b**). The data were been collected on every ninth ascending part of the orbit covering a wide latitude range from $-75^0$ south to $+75^0$ north at the time interval from May 5 to May 8, 2009, with a unique 2-seconds temporal resolution.

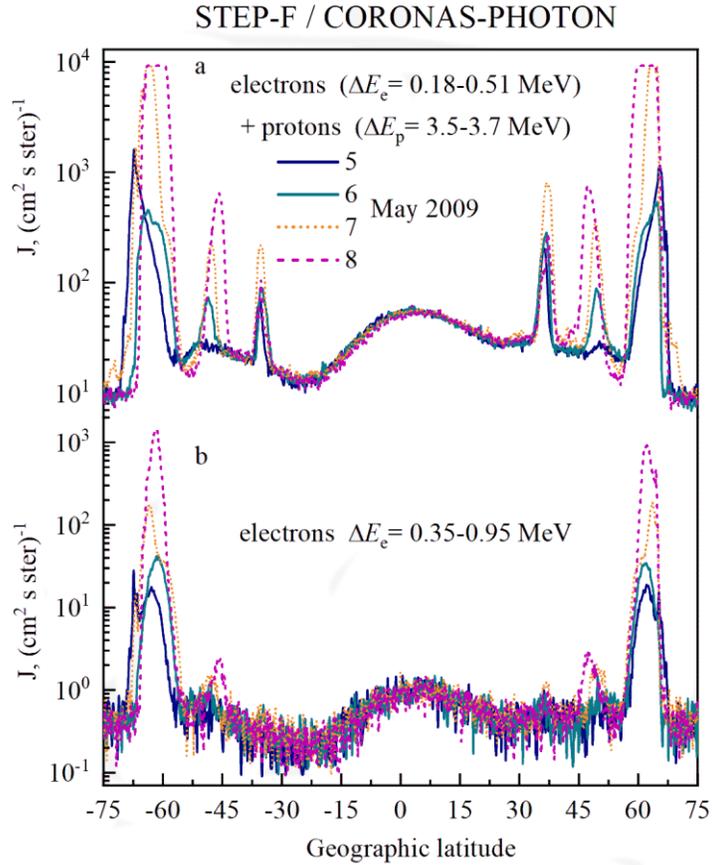

Figure 5. The fluence of particle intensity throughout the period of 5 to 8 May 2009 as measured by STEP-F at ascending part of the orbit. Different colors and line styles represent individual days as indicated.

In Figure 5 the satellite passage through the outer radiation belt in the southern and northern hemispheres is clearly seen as well as crossings of the Van Allen inner radiation belt in both hemispheres. Crossings through the third radiation belt identified in this work are observed in the southern and northern hemispheres at $\sim \pm 35^0$ latitudes. The presence of the third belt is recorded in both particle energy ranges but is much more pronounced at lower energies. An analysis of Figure 4 indicates that the main phase of the geomagnetic storm is associated with a sharp and substantial (more than one order of magnitude) increase of electron fluxes at the lower altitudes of the Van Allen outer belt in the two energy ranges of the STEP-F instrument, accompanied by a considerable enhancement of 2-MeV electron fluxes at the geostationary orbit as observed by the GOES-11 spacecraft. Plot "**e**" in Figure 4 shows a sharp increase and persistent electron fluxes on the geostationary orbit during the whole recovery phase of the geomagnetic storm up to May 14. The persistence of increased particle fluxes within the same time interval is also observed at the CORONAS-Photon circular orbit. Thus, there is a good correspondence between GOES and STEP-F/CORONAS-Photon data. Moreover, we have observed a definite correspondence of the particle intensity pulsations at various stages of the weak geomagnetic storm.

The specific feature of particle flux variability is that the third belt was observed clearly on May 5, precisely the day of a sharp enlargement of the high-speed solar wind density while the Van Allen belts particle fluxes did not differ from pre-storm levels with typical variations of tens of a percent because of the high sensitivity of the STEP-F instrument (see Figure 5, days May, 5,6). It was only during the initial and main phases of substorm particles flows were increased by an order of magnitude or more (see Figure 5, days of May, 7,8).

4. **Use of the SphinX spectrophotometer for detection of high energy charged particles**

*4.1. Sensitivity of SphinX detectors to mixed particle background*

As was pointed out in Section 2.1, the data from the 255th energy channels of SDet1 and SDet2 sensors were used for the present analysis. Figure 6 shows a map of count rates averaged over the entire measurement period from February 20 to November 29, 2009. Most of the detector events counted in the last energy channel are caused by charged particles, although there are also possible events due to solar X-ray photons with energies greater than 15 keV and spurious counts arising from regular resets of detector charge sensitive preamplifier. The map has been reconstructed only from those parts of the SphinX data when the instrument was in X-ray eclipse to eliminate "solar originating" photon counts in the highest energy channel. Preamplifier reset events have also been suppressed by applying appropriate correction of the background level. It was found that the rate of reset events strongly depends on detector temperature, and thus the signal from temperature sensor placed inside the detector package was used to determine a background level as well as the minimum orbital count rates within equatorial regions. Based on that information, the background was found to be varying with time from ~0.3 to ~3 counts per second. A subtraction of the reset background from the overall count rate in the highest energy channel of SDet2 improved the quality of the reconstructed map substantially, especially in the Van Allen belts and equatorial regions, where the relatively low particle rate was most affected by spurious reset events.

On the map one clearly sees areas of the SAA and the radiation belts in the Northern and Southern hemispheres, a pattern typical for the well-established distribution of high energy particles. However, in view of the fact that SphinX's detectors had not been intended for observations of particle content, their reaction to the presence of particles was unknown due to a lack of respective prelaunch calibrations against particles.

In order to find "a posteriori" the sensitivity of silicon PIN SDet1 and SDet2 detector signals to the presence of ambient energetic particles of various types and energies we performed comparison of the signals recorded by SDet1 and SDet2 with calibrated STEP-F recordings during the geomagnetic substorm on May 8, 2009. SDet1 and SDet2 responses on the spacecraft's crossing of radiation belts and SAA coincided with electron flux enhancements seen by the STEP-F's detectors. Counting rates recorded by SDet1 had values more than counting rates on SDet2 by a factor of ~ 5–10 at all times with the single exception of the SAA crossing: in this region counting rates detected by both sensors were similar. Closer to the geometric center of the SAA, the particle intensity detected with SDet2 became even greater than that in SDet1. This indicates substantially different sensitivity of SDet1 and SDet2 readouts to change of the primary particle sort and spectra, in particular in zones of the inner magnetosphere. It should be noted that SDet1 and SDet2 detectors were different in their inner construction.

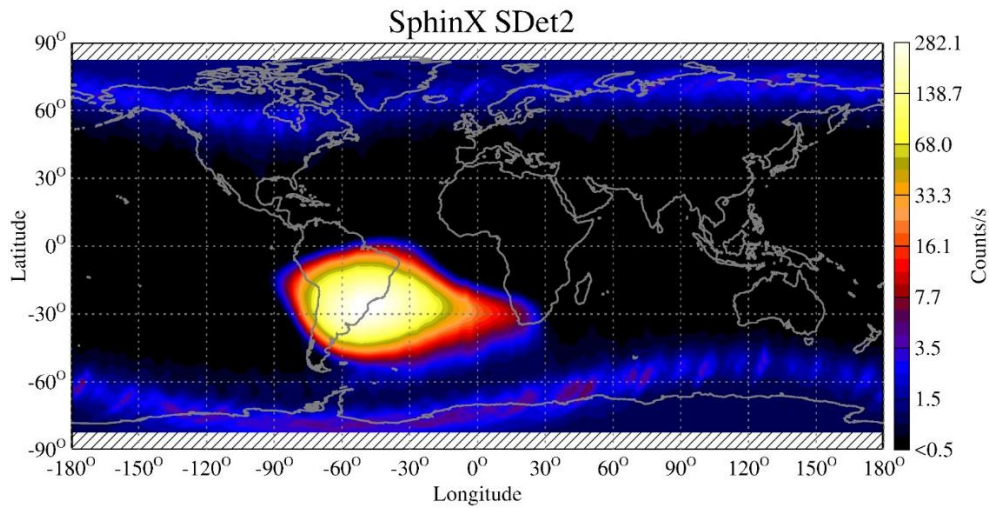

Figure 6. The geographical distribution of count rates seen by the SphinX SDet2 detector in its basic mode after applying a background level correction.

Detectors SDet1 and SDet2 were equipped with the front-end collimators limiting the field of view to a very narrow cone centered on the Sun. These collimators operated in the soft X-ray domain and for particles of sufficiently low energy. However, for high-energy gamma quanta and secondary emission arising from primary electron interactions with the detector head construction materials, the entire TESIS instrument [Kuzin et al., 2011] and the satellite itself, the emission can be observed from larger solid angles. By assuming that the electron energy spectrum outside the SAA is much softer than the one inside the SAA it is possible to explain the outstripping growth of count rates in SDet2 detector withing the SAA vs counts registered by SDet1 detector. This increase is due to the prevailing quantity of secondary gamma quanta generated as result of interaction of a highly energetic-portion of the primary electrons spectrum with the construction materials, and with the 400 μm tantalum plate covering SDet2 detector. This explains the observed disparity between ratio of their signals within and outside SAA. Therefore, SDet1 and SDet2 detectors of the SphinX instrument are sensitive to a mixture of electron ambient populations: electrons and the secondary γ–quanta, generated by magnetospheric high energy electrons.

The latter allowed us to introduce the concept of effective lowest energy thresholds for electron detection ($E_{thr1}$ and $E_{thr2}$ respectively) in SDet1 and SDet2 detectors, respectively. Corresponding values of these effective lowest threshold energies can slightly vary depending on the form of the electron energy spectrum.

It is worth noting the response of SphinX to protons is not thought to be significant. The arguments in favor are shown by [Li et al., 2013] where maps of electron and proton fluxes plotted vs geographic coordinates for the first 20 days of Relativistic Electron and Proton Telescope integrated little experiment (REPTile) operation onboard the Colorado Student Space Weather Experiment (CSSWE) CubeSat mission. The electrons in the [Li et al., 2013] Figure 5 are seen in a two-belt structure, with lower energy electrons penetrating into the lower $L$-region. Comparing the electron's map with our Figure 6 one can see a close similarity between both maps. On the contrary, energetic protons were detectable by REPTile when the CubeSat was above the SAA region only. The proton maps below differ noticeably from the map derived from SDtet2 data shown in Figure 6, so our conclusion is that protons don't contribute much to the SDet2 signal.

*4.2. The effective lowest threshold energies of electrons recorded by SphinX detectors inside the SAA*

To analyze the particle distributions concerning the position on the $L$–shells when the spacecraft crosses the SAA region, we analyzed ascending paths on every fifth orbit at each day in May 2009. In Figure 7 we present an example of derived particle flux density distributions for the four energy channels of the STEP–F instrument and for both SphinX detectors as recorded on May 11, 2009. The particle fluxes from SphinX detectors have not been normalized over solid angle. The dotted lines 7 and 8 in Figure 7 represent marks allowing for determination of $L$-values at which the largest particle fluxes have been recorded. It is seen that the ratio of counting rates between SDet1 and SDet2 detectors changes when the satellite crosses the SAA, varying its location between $L \sim 1.25$ and $L \sim 1.85$. A steeper decrease of SDet2 counting rates at higher $L$–shells indicates a significant change of the electron energy spectra related to a fast decrease of a high energy tail in the electron population. This is confirmed by the changing shape of the particle intensity profile recorded by the D1e detector between $L$–shells from $L \sim 1.5$ to $L \sim 1.9$, by a sharp decrease of electron fluxes recorded by the D2e and D1p detectors within referred above $L$-shells, and by the fast dropout of counting rates in the D4e detector, related to decrease of secondary γ – quanta.

The analysis of averaged data over the first fourteen days of May, 2009 $L$-shell values with highest particle intensities as a function of the electron energy (in our case energy channels, indicated in Table 1) allowed us to determine the values of effective lowest threshold energies $E_{thr1}$ and $E_{thr2}$ for the SphinX SDet1 and SDet2 detectors. It was also possible to determine the effective energy for the STEP–F $E_{D4e}$ detector as well. The Earth's magnetic field strength measured respective to the magnetic field strength on the geomagnetic equator during this period had values $B/B_0 \approx 1.2–1.6$. In Figure 8a we show the dependence of $L$–shells with measured largest particle fluxes on the electron energy-related to D2e and D1p channels in the period from 1 to 14 May 2009. This dependence within the concerned energy range is represented by a simple linear function:

$$L_{max}^{SAA} = \{1.44 \pm (2.7 \times 10^{-4})\} - \{0.35 \pm (5.5 \times 10^{-4})\} \times E \text{ (MeV)} \qquad (1)$$

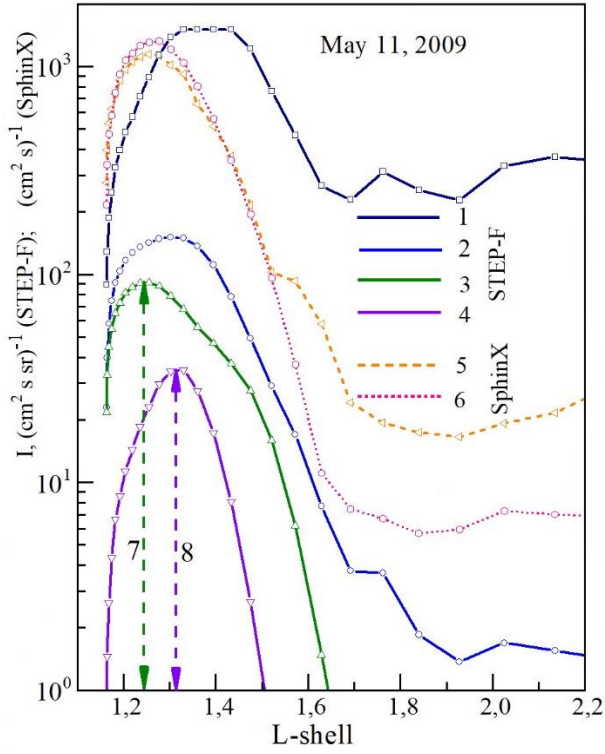

Figure 7. The particle intensity distribution recorded in energy channels D1e (**1**), D2e (**2**), D1p (**3**), D4e (**4**) of STEP-F, and SDet1 (**5**), SDet2 (**6**) of SphinX instruments.

D1e channel was not included because of the high-count rate in the SAA in this channel which exceeded the memory limit in the STEP-FE digital processing unit of the STEP-F instrument. That "saturation" is why it was also not possible to define the *L*-parameter value at which the largest particle flux density has been observed (see curve 1 in Figure 7).

The derived dependence (1) and knowledge of *L*-parameters at which the largest particle flux density was observed in the energy channel D4e of STEP-F (see dotted vertical line 7 in Figure 7) and in SDet1, SDet2 sensors of SphinX allow us to determine $E_{thr1}$, $E_{thr2}$, and $E_{D4e}$ values. From Figure 8a it is seen that the derived values of $E_{thr1}$ and $E_{thr2}$ are very close, assuming values ~ 500 keV and ~ 475 keV for the SDet1 and SDet2 sensors respectively. Slight uncertainty in these values is related to a poor statistic (14 days analyzed only), and also by discussed diurnal displacement of the satellite position along longitude at fixed latitude relative to the initial time. These effects impose slight day-to-day changes of the observed electron spectrum.

From Figure 8a the value of effective lowest threshold energy $E_{D4e}$ is observed to be ≈ 335 keV. The $E_{D4e}$ values varied slightly from day to day in the energy range between 230 to 350 keV during the period of May, 1 – 14, 2009. The lower values of the low threshold energy for electron detection in the detectors of the STEP-FD telescopic system as compared with respective $E_{thr1}$ and $E_{thr2}$ of the SphinX spectrophotometer is due to the high sensitivity of the STEP-F scintillation detector, with its large active area as well as the its very low noise of the photomultiplier tube electronics.

### 4.3. Parametrization of the $E_{thr1}$, $E_{thr2}$, and $E_{D4e}$ values as applied to the outer belt

As the electron energy spectra outside the SAA region have a much softer character than those inside the SAA, the values of $E_{thr1}$, $E_{thr2}$, and $E_{D4e}$ as applied to the outer radiation belt should differ from those ones determined inside the SAA. In order to estimate the $E_{thr1}$, $E_{thr2}$, and $E_{D4e}$ values outside the SAA we studied similarly the dependence of averaged values of *L*–parameter, for which maximum values of particle fluxes inside the outer belt were measured, with electron energy.

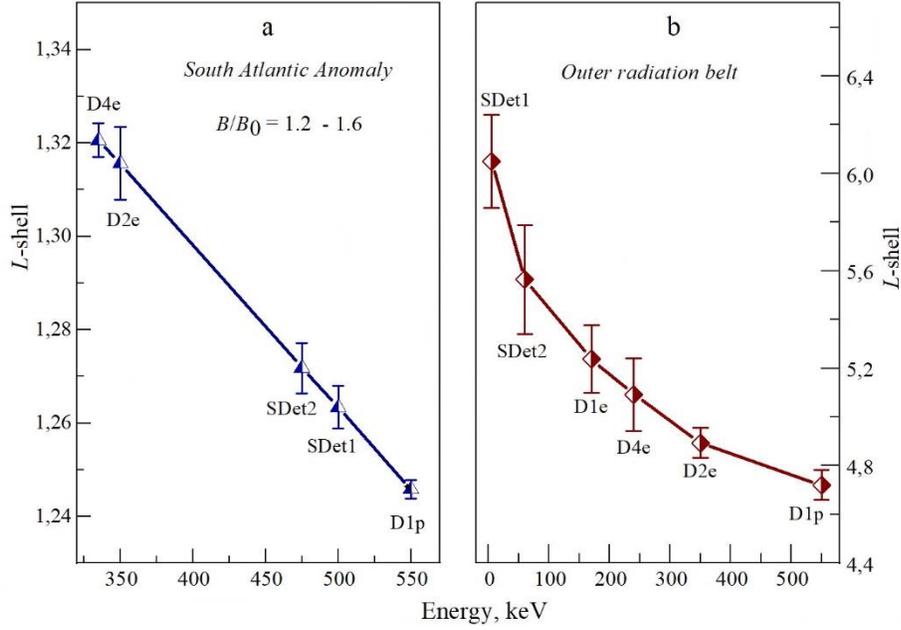

Figure 8. Dependences of averaged *L*-shell values, at which the largest electron fluxes were recorded inside the SAA (**a**) and in the outer belt in the Northern hemisphere (**b**), with particle energy.

The method relies on the fact that the higher particle energy is, the lower is its *L*-shell in the magnetosphere at which particles drift onto their way from one mirror point in the northern polar oval to the conjugate mirror point in the southern hemisphere [Horne et al., 2003; Lyons et al., 1972; Lyons, 1974].

The results shown in Figure 8b were derived for the ninth ascending part of the orbit in the northern hemisphere. The dependence of *L*–shell values, at which electrons reach greatest magnitude on the electron energy, can be approximated in the analyzed energy range as an exponential function:

$$L_{max}^{OutB} = (4.62 \pm 0.05) + (1.41 \pm 0.065) \times \exp\left(-\frac{E\ (MeV)}{0.21 \pm 0.02}\right) \qquad (2)$$

Taking this into account (see Figure 8b) the following values of a lowest effective threshold energies were determined: $E_{thr1} \approx 5$ keV, $E_{thr2} \approx 60$ keV, and $E_{D4e} \approx 240$ keV. The SphinX SDet2 detector was covered by the tantalum plate with a thickness of 400 μm as already mentioned - that is why it was unable to detect electrons with energies smaller than 1.2 MeV. However, the tantalum plate could serve as good target for the generation of low energy secondary gamma quanta as a result of its bombarding by the high energy magnetospheric electrons.

The effective low threshold energy $E_{thr2}$ value as applied to the outer radiation belt is smaller by a factor of few compared with that in the SAA region. This comes as a result of the softer character of electron spectrum in the radiation belt area. The larger is the number of high energy electrons in its spectrum the more flux of secondary emissions is being generated. In such a way the $E_{thr1}$ and $E_{thr2}$ values reflect the dominant presence or absence of a highly energetic portion of electrons. The $E_{thr2}$ is reduced also since a small part of the low energy electron fluxes enters detector window directly through the opening of area of $4.9 \times 10^{-3}$ cm$^2$ (used to measure the solar soft X-rays) and being subsequently recorded by a depletion layer of the large area PIN photodiode. The count rate of electrons recorded by the SDet1 sensor is higher than that for SDet2, but the contribution of low-energy γ–quanta to its total count is still significant.

Estimated values of the effective lowest threshold energy $E_{D4e}$ of D4e energy channel of the STEP-F instrument in the SAA and outer radiation belt regions do not differ significantly because of the specific logic mode incorporated in the electronic channel D4e of STEP-F operation. The registration of events in D4e is carried out in case the desired signals are observed simultaneously in all four layers of the detecting telescopic system. Meanwhile the decrease of the $E_{D4e}$ from ~ 335 keV in the SAA region to ~240 keV in the outer radiation belt also comes as a result of softening of the particle energy spectrum in the radiation belt, as compared with a slope of electron spectrum in the SAA.

5. **Evidence for the anisotropic electron fluxes in magnetospheric lower layers from interpretation of STEP-F and SphinX data**

Following the already described sensitivity calibration of SphinX detectors to particle fluxes, a cursory cross-examination of the data obtained by both STEP-F and SphinX indicates for presence of substantial spatial anisotropy in the electron fluxes in the radiation belts at the heights of ionosphere's F-layer. To verify this anisotropy hypothesis, we studied the time variation of the particle intensity during the period from 1$^{st}$ to 14$^{th}$ May, 2009 on every 13$^{th}$ descending part of the satellite orbit. The variability patterns were considered for the D1e and D2e detectors of the STEP-F, and for the SDet1 and SDet2 sensors of SphinX. Results are shown in Figure 9. The satellite crossed the outer radiation belt in the southern hemisphere at latitudes ≈ 80 to 72 degrees south, and in the northern hemisphere at latitudes ≈ 50 to 62 degrees (zone *1*); the inner radiation belt at the latitudes 62 to 50 degrees South, and at ≈ 35 to 40 degrees North (zone *2*); and the peripheral region of SAA -35 to -12 degrees South (zone *3*).

Figure 9 clearly indicates almost identical electron flux levels in both energy ranges of the STEP-F. At the same time the SphinX device did not detect any particle fluxes during the satellite passage through electron radiation belt in the northern hemisphere. However, the SDet1 sensor observed substantially enhanced count rates in the electron radiation belt of the southern hemisphere. The effect is also present in SDet2 sensor of the SphinX but to a smaller extent.

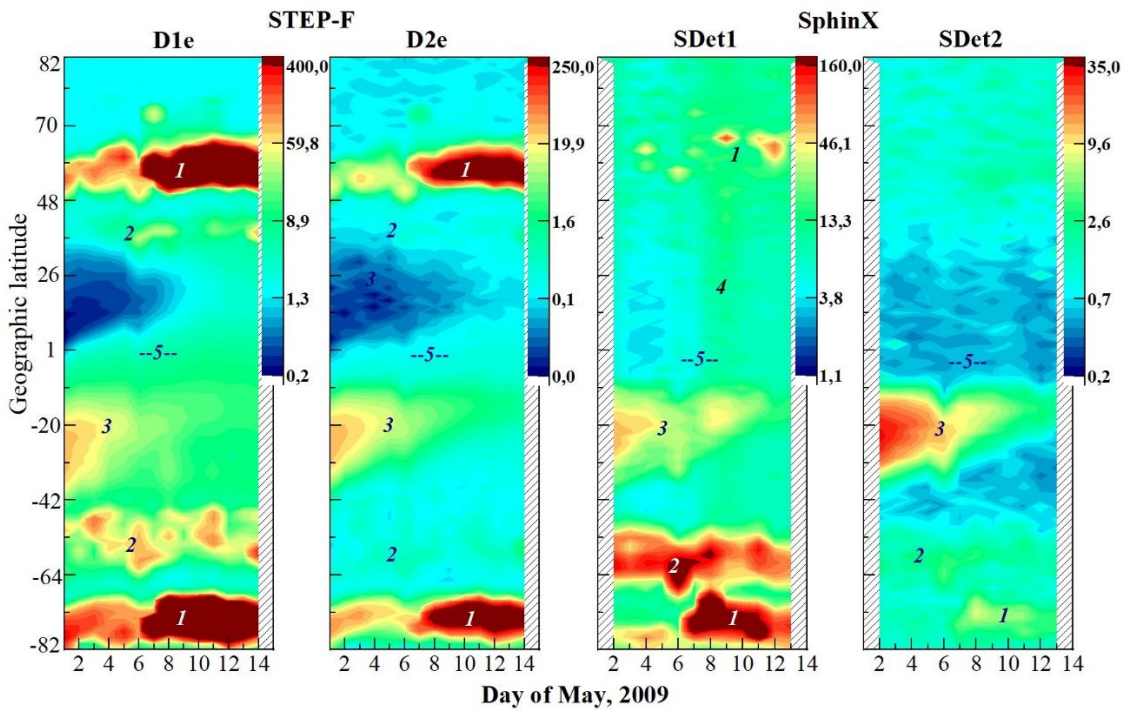

Figure 9. Patterns of electron intensity variations in the period from 1 to May 14, 2009 derived from D1e and D2e energy channels of the STEP-F, and from SDet1 and SDet2 sensors of the SphinX. The annotated regions on the plots are: *1*: Van Allen outer radiation belt; *2*: Van Allen inner radiation belt; *3*: peripheral region of the SAA; *4*: particles recorded by SDet1 sensor of SphinX during the main phase of the substorm; *5*: geographic equator.

Having in mind quite narrow field of view (< ~ 2°) of the SDet1 and SDet2 sensors in the SphinX device due to embedding collimators, such an asymmetry in the responses on satellite crossing of the same radiation belt in different hemispheres can be naturally explained by postulating presence of beams (strong directivity) for electron flows within the radiation belt. In other words, electron fluxes can be "on axis" of the SphinX collimators during crossing of the belt in the southern hemisphere and are off-axis during crossing of the same radiation belt in the northern hemisphere.

An essential distinction in SDet1 and SDet2 sensor count rates in the outer belt as recorded by SphinX in the southern hemisphere is the fact that they are not accompanied by corresponding differences in D1e and D2e channel readouts as recorded by STEP-F. A disparity is also observed for two more ratios, one of which is the ratio in count rates in the Van Allen outer and inner belts recorded by both, D1e and D2e channels, and the second one is a similar ratio recorded by a pair SDet1 and SDet2 sensors. The first distinction can be explained by the different slope in various parts of the particle energy spectrum. While SDet1 and SDet2 signals reflect presence of tens of keV particles, D1e and D2e signals reflect presence of particles at several hundreds of keV energy (see Figure 8a).

The observed effect may be explained by the preferred recording of secondary $\gamma$−radiation by both sensors of SphinX spectrophotometer. These $\gamma$−quanta have spectra steeply declining with increasing energy. Taking into account a very weak response of the D1 position-sensitive silicon matrix detector of STEP-F to the spacecraft crossing of the inner electron belt at the southern hemisphere, it can be concluded that the SDet1 silicon PIN photodiode of the SphinX device detected secondary $\gamma$-quanta generated by magnetospheric electrons of energies having upper limit defined by the STEP-F's D1e bottom threshold energy, i.e., $E_e < 180$ keV. This conclusion is also supported by the SDet1 sensor seeing precipitating electrons during the main phase of the geomagnetic substorm of May 8, 2009 over a wide range of latitudes (zone *4*) including the geographic equator (zone *5*).

It is worth noting that filling the gap between the radiation belts by energetic particles has been observed on many previous occasions [Baker et al., 2004; Baker and Kanekal, 2008; Hudson et al., 2008]. In our case due to rather narrow fields of view of both X-ray spectrophotometer's sensors, and because of the strongly directed character of particle fluxes in the radiation belts at the heights of ~ 550 km, enhanced particle fluxes in the gap between the radiation belts of the southern hemisphere was not detected because of a low magnitude of the substorm on May 8, 2009. A prominent region with the enhanced particle flows detected by the detector SDet2 sensor (right part of Figure 9) in the peripheral area of the SAA (zone *3*) shows evidence of a fundamental distinction in the spatial directivity and in the spectral shape of the energy distribution of electrons populating both the SAA and the Earth's radiation belts.

**5 Conclusions**

In the present study we analyzed variations and locations of magnetospheric regions filled with increased high energy electron fluxes beneath and outside the Van Allen radiation belts at a height of ~ 550 km during the 2009 minimum of the 11-year's cycle of the solar activity. The measurements for the analysis presented here were provided by two separate instruments, STEP-F and SphinX, placed next to each other aboard the CORONAS-Photon spacecraft. The larger geometric area and higher sensitivity of the STEP-F detectors allowed us to uncover the presence of additional, third persistent inner electron radiation belt. The signatures of this belt are clearly seen on a few diurnal parts of spacecraft orbit. The *L*-profile of this belt with the maximal intensity of electrons observed at the $L_{max} \approx 1.6$ resembles the *L*-profiles of the Van Allen belts, and does not represent intermittent-type events. The analysis of experimental data collected in the two energy ranges of STEP-F indicates a substantial difference between the energy spectra in the Van Allen inner belt and those at $L_{max} \approx 1.6$, i.e., harder spectrum being observed within the "classical" inner Van Allen belt. We did not detect electrons with energies $E_e \geq 0.4$ MeV in this discovered third electron radiation belt.

In order to characterize the sensitivity of the two X-ray detectors of the SphinX spectrophotometer using silicon PIN large area photodiodes to the presence of ambient energetic particles we have introduced the concept of an effective lowest energy thresholds of the electron detection. The particles contribute predominantly to the last two of the 256 energy channels. We defined and determined values of the lowest threshold energies $E_{thr1}$ for SDet1, and $E_{thr2}$ for SDet2 sensors of SphinX. We have found that the values of $E_{thr1}$ and $E_{thr2}$ within the SAA differ significantly from these for the outer radiation belt. This difference probably arises from differences in the energy spectra of electrons filling each of the regions. We point out that the SphinX measurements from two silicon sensors significantly widened and complemented the total energy range of the STEP-F measurements of particle energy spectra.

Finally, cross-comparison and back-to-back interpretation of data from the STEP-F and SphinX instruments for fourteen days of 2009 allowed us to show and discuss evidence for the anisotropic character of the electron distribution populating the lowest layers of the Earth's magnetosphere outside the SAA region. The peripheral area of the SAA detected by SDet2 sensor of the SphinX spectrophotometer supported the evidence for substantial differences in the energy spectra of electrons in the SAA and in the radiation belts.

## Acknowledgments

The authors acknowledge support from research project IRA NASU - SRC PAS in the frame of "Agreement on scientific cooperation between the Polish Academy of Sciences and the National Academy of Sciences of Ukraine".
The Polish co-authors acknowledge the support from the Polish National Science Centre grant: UMO-2020/39/B/ST9/01591.

## References

Arkhangelskaja, I. V., Amandjolova, D. B., Arhangelsky, A. I., and Yu. D. Kotov, 2008. Features of quasi-stationary precipitations according to the data obtained with the AVS-F instrument onboard the CORONAS-F satellite, Solar Syst. Res., 42(6), 536–542. https://doi.org/10.1134/S0038094608060063.

Baker, D. N., Kanekal, S. G., Li, X., Monk, S. P., Goldstein, J., and J. L. Burch, 2004. An extreme distortion of the Van Allen belt arising from the 'Halloween' solar storm in 2003, Nature, 432, 878–881. https://doi.org/10.1038/nature03116.

Baker, D. N., and S. G. Kanekal, 2008. Solar cycle changes, geomagnetic variations, and energetic particle properties in the inner magnetosphere, J. Atmos. Sol-Terr. Phys., 70(2-4), 195-206. https://doi.org/10.1016/j.jastp.2007.08.031.

Baker, D. N., Kanekal, S., Hoxie, V. C., Henderson, M. G., Li, X., Spence, H. E., Elkington, S. R., Friedel, R. H. W., Goldstein, J., Hudson, M. K., Reeves, G. D., Thotne, R. M., Kletzing, C. A., and S. G. Claudepierre, 2013. A Long-Lived Relativistic Electron Storage Ring Embedded in Earth's Outer Van Allen Belt, Science, 340(6129), 186-190. https://doi.org/10.1126/science.1233518.

Baker, D. N., Hoxie, V. C., Zhao, H., Jaynes, A. N., Kanekal, S. G., Li, X. and Elkington, S. R., 2019. Multiyear Measurements of Radiation Belt Electrons: Acceleration, Transport, and Loss. J. Geophy. Res., 124, 2588–2602. https://doi.org/10.1029/2018JA026259.

Blake, J. B., Kolasinski, W. A., Fillius, R. W., and E. G. Mullen, 1992. Injection of electrons and protons with energies of tens of MeV into L < 3 on 24 March 1991, Geophys. Res. Lett., 19(8), 821-824. https://doi.org/10.1029/92GL00624.

Blake, J. B., Looper, M. D., Baker, D. N., Nakamura, R., Klecker, B., and D. Hovestadt, 1996. New high temporal and spatial resolution measurements by SAMPEX of the precipitation of relativistic electrons, Adv. Space Res., 18(8), 171-186. https://doi.org/10.1016/0273-1177(95)00969-8.

Blake, J. B., and R. S. Selesnick, 2001. Studies of relativistic electron injection events in 1997 and 1998, J. Geophy. Res., 106(A9), 19157-19168. https://doi.org/10.1029/2000JA003039.

Blake, J. B., Inan U. S., Walt M., Bell, T. F., Bortnik, J., Chenette, D. L., and H. J. Christian, 2001. Lightning-induced energetic electron flux enhancements in the drift loss cone J. Geophys. Res., 106(A12), 29733-29744. https://doi.org/10.1029/2001JA000067.

Dudnik, O. V., Goka, T., Matsumoto, H., Fujii, M., Persikov, V. K., and T. V. Malykhina, 2003. Computer simulation and calibration of the charge particle spectrometer-telescope «STEP-F», Adv. Space Res., "Calibration, Characterization of Satellite Sensors, Physical Parameters Derived from Satellite Data", 32(11), 2367-2372. https://doi.org/10.1016/S0273-1177(03)90567-9.

Dudnik, O. V., 2010. Investigation of the Earth's radiation belts in May 2009 at the low orbit satellite with the STEP-F instrument (in Russian), ISSN 1561-8889 Space Science and Technology (Kosmichna Nauka i Tekhnolohiia), 16(5), 12-28. https://doi.org/10.15407/knit2010.05.012.

Dudnik, A. V., Persikov, V. K., Zalyubovsky, I. I., Timakova, T. G., Kurbatov, E. V., Kotov, Yu. D., and V. N. Yurov, 2011. High sensitivity STEP-F spectrometer–telescope for high-energy particles of the CORONAS-

PHOTON satellite experiment, Solar Syst. Res., 45(3), 212–220. https://doi.org/10.1134/S0038094611020043.

Dudnik, O., Sylwester, J., and P. Podgórski, 2014. Properties of magnetospheric high energy particles based on analysis of data from STEP-F and SphinX instruments onboard the "CORONAS-Photon" satellite, in "Space Research in Ukraine, 2012−2014. The Report to the COSPAR", ISBN 978-966-360-255-4, ed. by O.P.Fedorov, Kyiv: Publ. House "Akademperiodika", 53-61.

Dudnik, O. V., Sylwester, J., Kowaliński, M., Barylak, J., 2019. Utilization of design features of the particle telescope STEP-F and solar x-ray spectrophotometer SphinX for exploration of the Earth's radiation belt properties, Proc. SPIE, Vol. 11176 "Photonics Applications in Astronomy, Communications, Industry, and High-Energy Physics Experiments", 111763L-1 - 111763L-10. https://doi.org/10.1117/12.2537296.

Gburek, S., Siarkowski, M., Kepa, A., Sylwester, J., Kowalinski, M., Bakala, J., Podgorski, P., Kordylewski, Z., Plocieniak, S., Sylwester, B., Trzebinski, W., and S. Kuzin, 2011a. Soft X-ray variability over the Present Minimum of Solar Activity as Observed by SphinX, Sol. Syst. Res., 45(2), 182–187. https://doi.org/10.1134/S0038094611020055.

Gburek, S., Sylwester, J., Kowalinski, M., Bakala, J., Kordylewski, Z., Podgorski, P., Plocieniak, S., Siarkowski, M., Sylwester, B., Trzebinski, W., Kuzin, S. V., Pertsov, A. A., Kotov, Yu. D., Farnik, F., Reale, F., and K. J. H. Phillips, 2011b. SphinX soft X-ray spectrophotometer: science objectives, design and performance, Solar Syst. Res., 45(3), 189–199. https://doi.org/10.1134/S0038094611020067.

Grachev, E. A., Grigoryan, O. R., Klimov, S. I., Kudela, K., Petrov, A. N., Schwingenschuh, K., Sheveleva, V. N., and J. Stetiarova, 2005. Altitude distribution analysis of electron fluxes at L = 1.2–1.8, Adv. Space Res., 36(10), 1992-1996. https://doi.org/10.1016/j.asr.2003.03.078.

Grigoryan, O. R., Kudela, K., Rothkaehl, H., and V. Sheveleva, 2008a. The electron formations under the radiation belts at L-shells 1.2–1.9, Adv. Space Res., 41(1), 81-85. https://doi.org/10.1016/j.asr.2006.11.008.

Grigoryan, O. R., Panasyuk, M. I., Petrov, V. L., Sheveleva, V. N., and A. N. Petrov, 2008b. Spectral characteristics of electron fluxes at L < 2 under the radiation belts, Adv. Space Res., 42(9), 1523-1526. https://doi.org/10.1016/j.asr.2007.12.009.

Horne, R. B., Meredith, N. P., Thorne, R. M., Heynderickx, D., Iles, R. H. A., and R. R. Anderson, 2003. Evolution of energetic electron pitch angle distributions during storm time electron acceleration to megaelectronvolt energies, J. Geophys. Res., 108(A1, 1016). https://doi.org/10.1029/2001JA009165.

Hudson, M. K., Kress, B. T., Mueller, H.-R., Zastrow, J. A., and J. B. Blake, 2008. Relationship of the Van Allen radiation belts to solar wind drivers, J. Atmos. Sol-Terr. Phy., 70(5), 708-729. https://doi.org/10.1016/j.jastp.2007.11.003.

Imhof, W. L., Reagan, J. B., and E. E. Gaines, 1978. The energy selective precipitation of inner zone electrons, J. Geophys. Res., 83(A9), 4245-4254. https://doi.org/10.1029/JA083iA09p04245.

Imhof, W. L., Kilner, J. R., and J. B. Reagan, 1985. Morphological study of energetic electron precipitation events using the satellite bremsstrahlung X-ray technique, J. Geophys. Res., 90(A2), 1543-1552. https://doi.org/10.1029/JA090iA02p01543.

Imhof, W. L., Voss, H. D., Mobilia, J., Datlowe, D. W., Gaines, E. E., McGlennon, J. P., and U. S. Inan, 1992. Relativistic Electron Microbursts, J. Geophys. Res., 97(A9), 13829-13837. https://doi.org/10.1029/92JA01138.

Kotov, Yu. D., 2011. Scientific Goals and Observational Capabilities of the CORONAS-PHOTON Solar Satellite Project, Sol. Syst. Res., 45(2), 93–96. https://doi.org/10.1134/S0038094611020079.

Kudin, A. M., Borodenko, Yu. A., Grinyov, B. V., Didenko, A. V., Dudnik, A. V., Zaslavsky, B. G., Valtonen, E., Eronen, T., Peltonen, J., Lehti, J., Kettunen, H., Virtanen, A., and J. Huovelin, 2010. CsI(Tl) + photodiode scintillation assemblies for γ-ray and proton detectors, Instrum. Exper. Tech-U, 53(1), 39–44. https://doi.org/10.1134/S0020441210010057.

Kuzin, S. V., Zhitnik, I. A., Shestov, S. V., Bogachev, S. A., Bugaenko, O. I., Ignat'ev, A. P., Pertsov, A. A., Ulyanov, A. S., Reva, A. A., Slemzin, V. A., Sukhodrev, N. K., Ivanov, Yu. S., Goncharov, L. A., Mitrofanov, A. V., Popov, S. G., Shergina, T. A., Solov'ev, V. A., Oparin, S. N., Zykov, A. M., 2011. The TESIS experiment on the CORONAS-PHOTON spacecraft, Solar Syst. Res., 45(2), 162–173. https://doi.org/10.1134/S0038094611020110.

Li, X., Schiller, Q., Blum, L., Califf, S., Zhao, H., Tu, W., Turner, D. L., Gerhardt, D., Palo, S., Kanekal, S., Baker, D. N., Fennel, J., Blake, J. B., Looper, M., Reeves, G. D., Spence, H., 2013. First results from CSSWE CubeSat: Characteristics of relativistic electrons in the near-Earth environment during the October 2012 magnetic storms. J. Geophys. Res., 118, 6489–6499. https://doi.org/10.1002/2013JA019342.

Lorentzen, K. R., Looper, M. D., and J. B. Blake, 2001. Relativistic electron microbursts during the GEM storms, Geophys. Res. Lett., 28(13), 2573-2576. https://doi.org/10.1029/2001GL012926.

Lyons, L. R., Thorne, R. M., and C. F. Kennel, 1972. Pitch-angle diffusion of radiation belt electrons within the plasmasphere, J. Geophys. Res., 77, 3455–3474. https://doi.org/10.1029/JA077i019p03455.

Lyons, L. R., 1974. Pitch angle and energy diffusion coefficients from resonant interactions with ion-cyclotron and whistler waves, J. Plasma Phys., 12(3), 417-432. https://doi.org/10.1017/S002237780002537X.

Martinez-Calderon, C., Bortnik, J., Li, W., Spence, H. E., Claudepierre, S. G., Douma, E., and C. J. Rodger, 2020. Comparison of long-term lightning activity and inner radiation belt electron flux perturbations, J. Geophys. Res., 125, e2019JA027763. https://doi.org/10.1029/2019JA027763.

Millan, R. M., and the BARREL team, 2011. Understanding relativistic electron losses with BARREL, J. Atmos. Sol-Terr. Phys., 73(11-12), 1425-1434. https://doi.org/10.1016/j.jastp.2011.01.006.

Millan, R. M., McCarthy, M. P., Sample, J. G., Smith, D. M., Thompson, L. D., et al., 2013. The Balloon Array for RBSP Relativistic Electron Losses (BARREL), Space Sci. Rev., 179(1-4), 503-530. https://doi.org/10.1007/s11214-013-9971-z.

Nagata, K., Kohno, T., Murakami, H., Nakamoto, A., Hasebe, N., Kikuchi, J., and T. Doke, 1988. Electron (0.19-3.2 MeV) and proton (0.58-35 MeV) precipitations observed by ONZORA satellite at low latitude zones L=1.6-1.8, Planet. Space Sci., 36(6), 591-606. https://doi.org/10.1016/0032-0633(88)90028-1.

Nakamura, R., Baker, D. N., Blake, J. B., Kanekal, S., B. Klecker, and D. Hovestadt, 1995. Relativistic electron precipitation enhancements near the outer edge of the radiation belt, Geophys. Res. Lett., 22(9), 1129-1132. https://doi.org/10.1029/95GL00378.

Nakamura, R., Isowa, M., Kamide, Y., Baker, D. N., Blake, J. B., and M. Looper, 2000. SAMPEX observations of precipitation bursts in the outer radiation belt, J. Geophys. Res., 105(A7), 15875-15885. https://doi.org/10.1029/2000JA900018.

Obara, T., Matsumoto, H., Koga, K., 2012. Space environment measurements by JAXA satellite and ISS/JEM, Acta Astronaut., 71, 1-10. https://doi.org/10.1016/j.actaastro.2011.08.009.

Rodger, C. J., Clilverd, M. A., and McCormick, R. J., 2003. Significance of lightning-generated whistlers to inner radiation belt electron lifetimes. J. Geophys. Res., 108(A12), 1462. https://doi.org/10.1029/2003JA009906.

Sadovnichy, V. A., Panasyuk, M. I., Yashin, I. V., Barinova, V. O., Veden'kin, N. N., et. al. Investigations of the space environment aboard the Universitetsky–Tat'yana and Universitetsky–Tat'yana-2 microsatellites, 2011. Solar Syst. Res., 45(1), 3–29. https://doi.org/10.1134/S0038094611010096.

Sylwester, J., Kuzin, S., Kotov, Yu. D., Farnik, F., and F. Reale, 2008. SphinX: A fast solar Photometer in X-rays, J. Astrophys. Astron., 29(1-2), 339-343. https://doi.org/10.1007/s12036-008-0044-8.

Woodger, L. A., Halford, A. J., Millan, R. M., McCarthy, M. P., Smith, D. M., Bowers, G. S., Sample, J. G., Anderson, B. R., and Liang, X., 2015. A summary of the BARREL campaigns: Technique for studying electron precipitation, J. Geophys. Res. Space Physics, 120, 4922–4935. http://doi.org/10.1002/2014JA020874.

Van Allen, J. A., Ludwig, G. H., Ray, E. C., and C. E. McIlwain, 1958. Observations of high intensity radiation by satellites 1958 Alpha and Gamma / J. Jet Propulsion, 28(9), 588-592. https://doi.org/10.2514/8.7396.

Vernov, S. N., Grigorov, N. L., Ivanenko, I. D., Lebedinskii, A. I., Murzin, V. S., and A. E. Chudakov, 1959. Possible mechanism of production of terrestrial corpuscular radiation under the action of cosmic rays, Reports of Academy of Sciences USSR, 124(5), 1022-1025.